\documentclass[
    ,final            
   ,numberedheadings 
  ]
  {aipproc}

\layoutstyle{8x11single}

\begin{document}

\title{The Relativistic Inverse Stellar Structure Problem}


\author{Lee Lindblom}{
  address={Theoretical Astrophysics, California Institute of Technology, 
 Pasadena, CA 91125}
}

\begin{abstract}
  The observable macroscopic properties of relativistic stars (whose
  equations of state are known) can be predicted by solving the
  stellar structure equations that follow from Einstein's equation.
  For neutron stars, however, our knowledge of the equation of state
  is poor, so the direct stellar structure problem can not be solved
  without modeling the highest density part of the equation of state
  in some way.  This talk will describe recent work on developing a
  model independent approach to determining the high-density
  neutron-star equation of state by solving an inverse stellar
  structure problem.  This method uses the fact that Einstein's
  equation provides a deterministic relationship between the equation
  of state and the macroscopic observables of the stars which are
  composed of that material.  This talk illustrates how this method
  will be able to determine the high-density part of the neutron-star
  equation of state with few percent accuracy when high quality
  measurements of the masses and radii of just two or three neutron
  stars become available.  This talk will also show that this method
  can be used with measurements of other macroscopic observables, like
  the masses and tidal deformabilities, which can (in principle) be
  measured by gravitational wave observations of binary neutron-star
  mergers.
\end{abstract}

\maketitle 
 

\section{Standard Stellar Structure Problem}

Models of stars in general relativity theory are constructed by
solving the relativistic stellar structure equations that follow from
Einstein's equation.  The most widely used form of these equations,
for non-rotating stellar models, was first derived by Oppenheimer and
Volkoff~\cite{Oppenheimer1939}:
\begin{eqnarray}
\frac{dm}{dr} &=& 4\pi r^{\,2} \epsilon,
\label{e:OVmass}\\
\frac{dp}{dr} &=& -(\epsilon+p)\frac{m + 4\pi r^{\,3} p}{r(r-2m)}.
\label{e:OVpressure}
\end{eqnarray}
These equations determine $m(r)$ and $p(r)$, the mass of the star
contained within a sphere of radius $r$ and the pressure respectively.
These equations are incomplete however, and must be supplemented by
giving an equation of state that characterizes the material contained
within the star.  In particular, the relationship between the total
energy density $\epsilon$ and the pressure $p$ must be specified.  The
addition of the relationship $\epsilon=\epsilon(p)$ completes the
structure equations, (\ref{e:OVmass}) and (\ref{e:OVpressure}), making
them a first-order set of ordinary differential equations that can be
solved using standard techniques.  These equations are usually
integrated starting at the center of the star, where $r=0$, using the
boundary conditions $m(0)=0$, to ensure that the stellar model is
non-singular, and $p(0)=p_c$ to select the central value of the
pressure.  The value of the central pressure $p_c$ can be chosen
arbitrarily, with larger $p_c$ generally corresponding to stellar
models having larger masses.  The stellar structure equations,
(\ref{e:OVmass}) and (\ref{e:OVpressure}), are then integrated toward
larger values of $r$ until the surface of the star at $r=R$ is
reached, where $p(R)=0$.  The total mass of the star is given by
$M=m(R)$.

\begin{figure}[!t]
\includegraphics[width=4.4in]{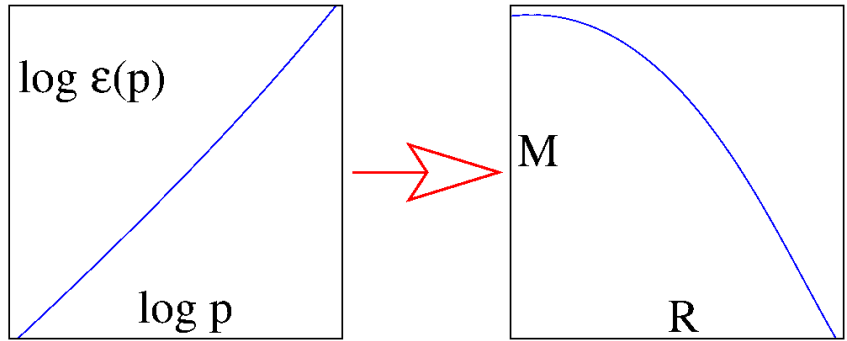}
\caption{Standard relativistic stellar structure problem can be
  thought of as an abstract map between an equation of state (a curve
  in the space of densities and pressures) and a curve of macroscopic
  observables, like the mass-radius curve.
\label{f:Fig1}}
\end{figure}

The central pressure $p_c$ is a freely specifiable parameter, so the
stellar structure equations determine a one-parameter family of
solutions, $m(r,p_c)$ and $p(r,p_c)$, for any given equation of state.
In general, the macroscopic observables associated with these stellar
models, like the total mass $M$ and radius $R$, also depend on this
parameter $p_c$.  Therefore the macroscopic observables associated
with any particular equation of state can be thought of as a curve,
parameterized by $p_c$, in the space of those observables.  This talk
will focus on the particular case of masses $M(p_c)$ and radii
$R(p_c)$.  The standard stellar structure problem can be thought of as
an abstract map that associates an equation of state (which is in
effect a curve though the space of energy densities and pressures)
with a curve through the space of observables.  Figure~\ref{f:Fig1}
illustrates what this abstract map looks like for the particular case
of the mass-radius curve of observables.

\section{Inverse Stellar Structure Problem}

\begin{figure}[!hb]
\includegraphics[width=2.0in]{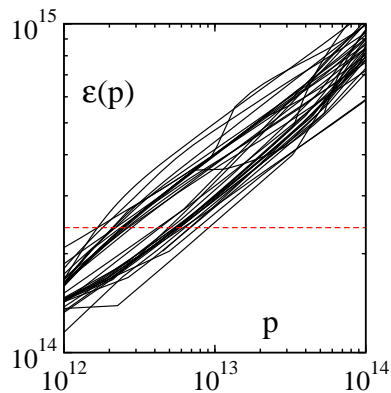}
\caption{The neutron-star equation of state is poorly known, as
  illustrated here by the high-density portions of 34 theoretical
  neutron-star equations of state.  The range of pressures at nuclear
  density (dashed red line) in these equations of state span about an
  order of magnitude. \label{f:Fig2}}
\end{figure}

When the equation of state is well understood, as in white dwarf
stars, the standard stellar structure problem is very useful.
Detailed models of these stars can be computed, and their observable
properties can be compared directly with observations.  For cases
where the equation of state is poorly know however, like neutron
stars, the standard stellar structure problem is not so useful.
Figure~\ref{f:Fig2} illustrates the high-density portions of 34
theoretical neutron-star equations of state~\cite{Read:2008iy}.  The
range of pressures corresponding to nuclear density (dashed red line)
predicted by these theoretical equations of state have a range that
spans about an order of magnitude.  This means that it is not possible
to make reliable quantitative predictions about the structures of
neutron stars simply by solving the standard stellar structure
problem.

The range of densities and pressures inside typical neutron stars are
well beyond the range where any current or foreseeable future
laboratory experiment will be able to measure the neutron-star
equation of state directly.  In contrast, observations of the
macroscopic properties of neutron stars are becoming more numerous and
more accurate all the time.  What is needed then is an inverse stellar
structure problem, where the equation of state is determined from a
knowledge of the macroscopic observables of those stars, e.g. their
masses and radii.  That is, we need a way of solving the stellar
structure equations for the equation of state, instead of specifying
it {\sl a priori} as we did as part of the definition of the standard
stellar structure problem.

\begin{figure}[!bth]
\includegraphics[width=4.4in]{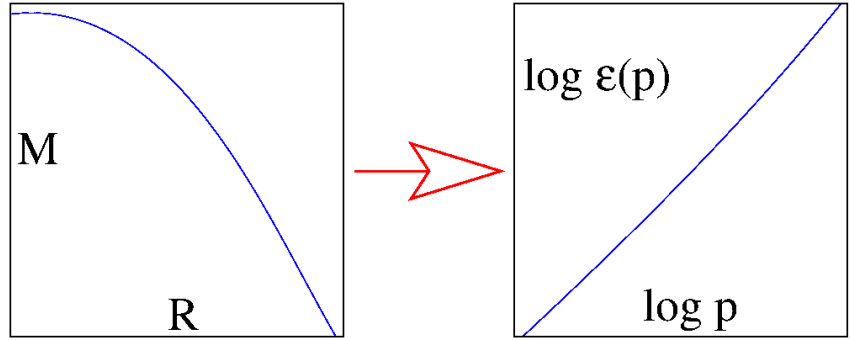}
\caption{Inverse stellar structure problem can be thought of as an
  abstract map that takes a curve of observables, like the mass-radius
  curve, and from it determines the equation of state of the stellar
  matter.\label{f:Fig3}}
\end{figure}

At an abstract level it is not hard to imagine what such an inverse
stellar structure problem might look like.  In Fig.~\ref{f:Fig1} we
visualized the standard stellar structure problem as a map that takes
a curve in density-pressure space (i.e. the equation of state) and
identifies it with a curve in the space of macroscopic observables.
From this perspective, the inverse stellar structure problem
corresponds to the inverse of this abstract map.  That is, the inverse
stellar structure problem is the abstract map, illustrated in
Fig.~\ref{f:Fig3}, that takes a curve in the space of macroscopic
observables and from it determines the corresponding curve in
density-pressure space.

\section{Traditional Solution of the Inverse Stellar Structure Problem}

Let me first discuss a particular solution to the inverse stellar
structure problem, developed about 20 years ago~\cite{Lindblom1992},
that I call the traditional solution.  Later in this talk I will also
discuss a newer and more sophisticated approach to solving this
problem.  The traditional approach, however, provides a clearer
insight into the way the stellar structure equations determine the
solution to this inverse problem, so I think it is worth discussing.

\begin{figure}[!hb]
\includegraphics[width=6.5in]{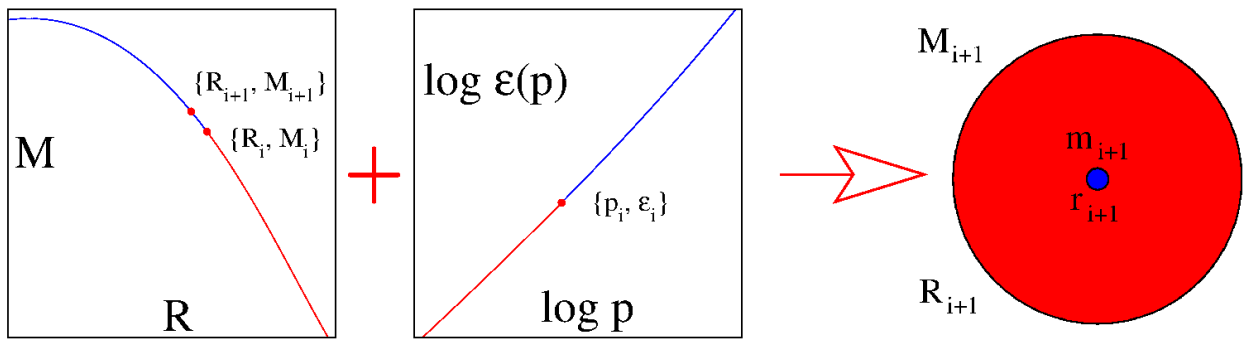}
\caption{Knowledge of the complete mass-radius curve plus a knowledge
  of the low density equation of state determines the structure of the
  outer parts of the star with total mass $M_{i+1}$ and radius
  $R_{i+1}$.  Also determined from this knowledge are the mass
  $m_{i+1}$ and radius $r_{i+1}$ of the small inner core of material
  having densities and pressures from the undetermined high-density
  part of the equation of state.\label{f:Fig4}}
\end{figure}

Let us assume that the entire mass-radius curve $\{R,M\}$ is known.
Let us also assume that the lowest density part of the equation of
state is known, i.e., $\epsilon=\epsilon(p)$ is assumed to be known
for $p\leq p_i$ and $\epsilon(p)\leq \epsilon(p_i)=\epsilon_i$.  Let
$\{R_i,M_i\}$ denote the point on the mass-radius curve whose central
pressure is $p_i$.  We choose another point on the mass-radius curve,
$\{R_{i+1},M_{i+1}\}$, that is ``close'' to the point $\{R_i,M_i\}$,
but has a slightly larger central pressure.  Figure~\ref{f:Fig4}
illustrates these assumptions.  The mass-radius curve is illustrated
in the first panel of Fig.~\ref{f:Fig4}, including the particular
points $\{R_i,M_i\}$ and $\{R_{i+1},M_{i+1}\}$. The red portion of the
equation of state curve, shown in the middle panel of
Fig.~\ref{f:Fig4}, is also assumed to be known, including the
particular point $\{p_i,\epsilon_i\}$.  The red portion of the
mass-radius curve in Fig.~\ref{f:Fig4} represents those stellar models
whose central pressures are less than $p_i$.  The panel on the right
in Fig.~\ref{f:Fig4} is a cross section that represents the structure
of the stellar model $\{R_{i+1},M_{i+1}\}$.  The outer portions of
this model, shaded red, are composed of low density material whose
equation of state is already known.  The small central core of this model,
shaded blue, is composed of material whose pressure and density exceed
$\{p_i,\epsilon_i\}$.

The relativistic stellar structure equations, (\ref{e:OVmass}) and
(\ref{e:OVpressure}), can be integrated to determine quantitatively
the structure of the stellar model $\{R_{i+1},M_{i+1}\}$.  Generally
these equations are integrated from the center of the star, $r=0$,
outward.  But we can not do that in this case, because we do not know
the equation of state at densities contained in the core of this
model.  We can, however, integrate the equations starting at the
surface of this star, $r=R_{i+1}$, using as initial data
$m(R_{i+1})=M_{i+1}$ and $p(R_{i+1})=0$.  Since the low density
equation of state is known, this integration can be continued up to
the point where $p=p_i$.  The mass and radius, $\{r_{i+1},m_{i+1}\}$,
of this small high-density core can therefore be determined
quantitatively from this model: $p(r_{i+1})=p_i$, and
$m(r_{i+1})=m_{i+1}$.

\begin{figure}[!h]
\includegraphics[width=4.4in]{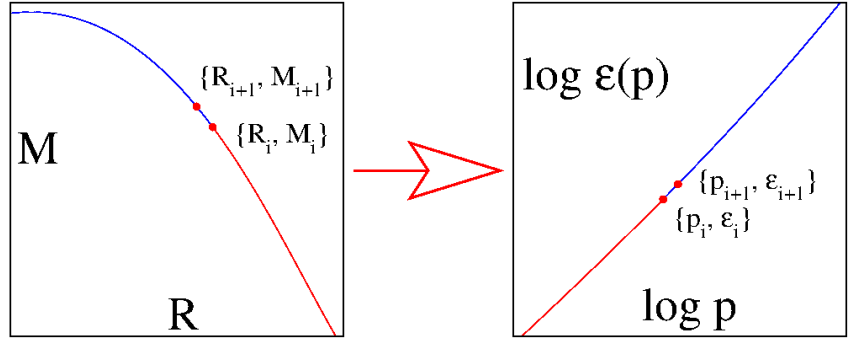}
\caption{Traditional solution of the inverse stellar structure problem
  iteratively determines a sequence of points on the equation of
state curve by inverting the power series solutions for small inner
cores of material whose masses and radii are determined numerically
from a knowledge of the complete mass radius curve.
\label{f:Fig5}}
\end{figure}

If the point $\{R_{i+1},M_{i+1}\}$ has been chosen to be sufficiently
close to $\{R_{i},M_{i}\}$, then the high-density core composed of
material from the unknown high-density part of the equation of state
will have very small mass and radius, $\{r_{i+1},m_{i+1}\}$. In this
case the structure equations, (\ref{e:OVmass}) and
(\ref{e:OVpressure}), can be solved as power series expansions:
\begin{eqnarray}
m_{i+1}&=&\frac{4\pi}{3}\epsilon_{i+1}r_{i+1}^3+{\mathcal O}(r_{i+1}^5),
\label{e:OVmassSeries}\\
p_i&=&p_{i+1}-\frac{2\pi}{3}(\epsilon_{i+1}+p_{i+1})(\epsilon_{i+1}+3p_{i+1})r_{i+1}^2
+{\mathcal O}(r_{i+1}^4).\label{e:OVpressureSeries}
\end{eqnarray}
The coefficients in these series depend on
$\{p_{i+1},\epsilon_{i+1}\}$, the central density and pressure of the
little core $\{r_{i+1},m_{i+1}\}$.  These series can therefore be
inverted to determine $\{p_{i+1},\epsilon_{i+1}\}$ in terms of the
known quantities $\{p_{i},\epsilon_{i}\}$ and $\{r_{i+1},m_{i+1}\}$.
This provides an additional point on the equation of state curve, as
illustrated in Fig.~\ref{f:Fig5}.  The points on the equation of state
curve that are intermediate between $\{p_{i},\epsilon_{i}\}$ and
$\{p_{i+1},\epsilon_{i+1}\}$ are determined by adopting a suitable
interpolation formula.  This procedure can be iterated to determine
the equation of state for the entire range of densities that occur
within the cores of neutron stars.  This method has been used
successfully to solve the inverse stellar structure problem using mock
mass radius data constructed from a known theoretical equation of
state in Lindblom~\cite{Lindblom1992}.

This traditional method for solving the inverse stellar structure
problem represents the equation of state as a table
$\{p_i,\epsilon_i\}$ for $i=1,...,N_{EOS}$ plus an
interpolation formula to fill in intermediate points.  The accuracy of
a particular implementation of this method depends on how many terms
are retained in the power series expansions, (\ref{e:OVmassSeries})
and (\ref{e:OVpressureSeries}), as well as the type of interpolation
formula used.  The overall accuracy of the equation of state
determined in this way will also depend on the number,
$N_\mathrm{stars}$, of mass-radius data available, typically scaling
like $N_\mathrm{stars}^{-n}$ for some $n$.  Since this method of
representing equations of state is rather inefficient, and since this
method of solving the inverse stellar structure problem converges
rather slowly, many mass-radius data points would be needed to
determine the equation of state with reasonable accuracy in this way.
Measurements of these data are very difficult to make, so it seems
unlikely that this traditional method will ever be useful for
analyzing real astrophysical data.
 
\section{Spectral Approach to the Inverse Stellar Structure Problem}

The key idea of the spectral approach to the inverse stellar structure
problem is to find a more efficient way to represent equations of
state using spectral methods.  Fewer observations of the macroscopic
observables should then be sufficient to determine the equation of
state by solving an appropriately modified version of the inverse
stellar structure problem.

The basic idea is to construct efficient parametric representations of
the equation of state, e.g. $\epsilon=\epsilon(p,\gamma_{\,k})$, where
the $\gamma_{\,k}$ are a set of parameters that fix the functional
form of a particular equation of state.  One example of such a
representation is a simple spectral expansion of the form,
$\epsilon(p,\gamma_{\,k})=\sum_k\gamma_{\,k}\Phi_k(p)$, where the
$\Phi_k(p)$ are suitably chosen basis functions.  Examples of such
basis functions include the Fourier basis functions
$\Phi_k(p)=e^{ikp}$, and the various collections of orthonormal
polynomials, like the Chebyshev polynomials $\Phi_k(p)=T_k(p)$.
Spectral expansions of this type are exact when an infinite number of
basis functions are included in the sum.  Truncating these expansions
at some finite number of terms, $k\leq N_{\gamma_{\,k}}$, gives an
approximation to the equation of state, whose errors decrease faster
than any power of $N_{\gamma_{\,k}}$ (e.g. typically exponentially)
for smooth functions.  I will discuss the particular choice of
parametric representation that has been useful for equations of state
later in this talk.

For now, let us simply assume that we have a parametric
representation, $\epsilon=\epsilon(p,\gamma_{\,k})$, that provides
reasonably accurate approximate equations of state with just a few
parameters $\gamma_{\,k}$.  We use this family of equations of state
to solve the standard stellar structure equations, (\ref{e:OVmass})
and (\ref{e:OVpressure}).  The results include the various macroscopic
observables associated with these stellar models, like their total
masses $M(p_c,\gamma_{\,k})$ and radii $R(p_c,\gamma_{\,k})$.  These
model observables now depend explicitly in the parameters
$\gamma_{\,k}$ that determine the particular equation of state, as well
as the parameter $p_c$ that specifies the central pressure of the
model.  The spectral approach to the inverse stellar structure problem
fixes these equation of state parameters, $\gamma_{\,k}$, by matching
the model observables $\{R(p_c,\gamma_{\,k}),M(p_c,\gamma_{\,k})\}$ to
a collection of points $\{R_i,M_i\}$, for $i=1,...,N_\mathrm{stars}$,
from the exact curve of macroscopic observables for these stars.  The
parameters $\gamma_{\,k}$, along with the parameters $p_c^{\,i}$ that
represent the central pressures of the models whose observables are
$\{R_i,M_i\}$, are determined by minimizing the quantity
$\chi(\gamma_{\,k},p_c^{\,i})$, defined by,
\begin{eqnarray}
\chi^2(\gamma_{\,k},p_c^{\,i}) &=& \frac{1}{N_{\,\mathrm{stars}}}
\sum_{i=1}^{N_{\,\mathrm{stars}}} \left\{
\left[\log\left(\frac{R(p_c^{\,i},\gamma_{\,k})}{R_i}\right)\right]^2
+\left[\log\left(\frac{M(p_c^{\,i},\gamma_{\,k})}{M_i}\right)\right]^2
\right\}, 
\end{eqnarray}
with respect to each of the parameters $\gamma_{\,k}$ and $p_c^{\,i}$
This non-linear least squares problem can be solved using standard
numerical methods.  Once fixed in this way, the parameters
$\gamma_{\,k}$ now determine an equation of state,
$\epsilon=\epsilon(p,\gamma_{\,k})$, that provides an approximate
solution to the inverse stellar structure problem.  The accuracies of
spectral expansions typically decrease exponentially with the number
of parameters $N_{\gamma_{\,k}}$ used in the expansion.  The
expectation is that the approximate equations of state constructed
with this spectral approach to the inverse stellar structure problem
should therefore converge rapidly to the real equation of state as the
number of parameters $\gamma_{\,k}$ is increased.

\section{Spectral Representations of Neutron-Star Equations of State}

The standard Oppenheimer-Volkoff representation of the stellar
structure equations, (\ref{e:OVmass}) and (\ref{e:OVpressure}), which
use the pressure as the fundamental thermodynamic variable, is not
actually the best form to use for numerical solutions
(cf. Lindblom~\cite{Lindblom1992}).  For a variety of reasons it is
better to re-write those equations in terms of the relativistic
enthalpy $h$, defined by
\begin{eqnarray}
h(p)&=&\int_0^p\frac{dp'}{\epsilon(p')+p'},
\label{e:hdef}
\end{eqnarray}
instead of the pressure $p$.  I will discuss some of the benefits of
the enthalpy-based stellar structure equations later in this talk.
But let us focus first on the problem of finding suitable parametric
enthalpy-based representations of the equation of state.  The
enthalpy-based equation of state consists of the pair of functions,
$\epsilon=\epsilon(h)$ and $p=p(h)$.  The simple spectral expansions
of these equations would have the forms
$\epsilon(h,\alpha_{\,k})=\sum_k\alpha_{\,k}\Phi_k(h)$ and
$p(h,\beta_{\,k})=\sum_k\beta_{\,k}\Phi_k(h)$.  These simple
expansions have two serious drawbacks.  First, the spectral basis
functions $\Phi_k(h)$ are typically oscillatory, while physical
equations of state must be monotonically increasing functions.  While
it is possible to express every monotonic function using an expansion
of this sort, doing so requires that the spectral coefficients satisfy
a very complicated set of inequalities.  The second drawback is that
these expansions of $\epsilon(h)$ and $p(h)$ require two sets of
parameters, while the original $\epsilon(p)$ expansion considered
previously required only one.  These simple enthalpy based expansions
are therefore very inefficient, which makes them inappropriate for our
purposes.

We need  {\sl faithful} spectral representations of the
equation of state.  Faithful representations in this context means
ones where every choice of spectral parameters corresponds to a
possible physical equation of state, and conversely ones where every
physical equation of state can be represented by some choice of
spectral parameters (cf. Lindblom~\cite{Lindblom10}).  The simple
spectral expansions proposed above are not faithful. It is not
possible to represent the space of all monotonically increasing
functions as all possible linear combinations of any choice of basis
functions, because monotonic functions do not form a vector space.
Faithful spectral expansions can, however, be constructed for the
adiabatic index, $\Gamma(h)$, defined as
\begin{eqnarray}
\Gamma(h) = \frac{\epsilon +p}{p}\frac{dp}{d\epsilon}
=\exp\left[\sum_k \gamma_{\,k} \Phi_k(h)\right].
\label{e:Gammadef}
\end{eqnarray}
Monotonicity of the equation of state implies that the adiabatic index
is positive, $\Gamma(h)>0$, but there is no requirement that it must
also be monotonic.  Any spectral expansion of the form given in
equation (\ref{e:Gammadef}) is therefore faithful.

Unfortunately, the adiabatic index $\Gamma(h)$ does not appear in the
stellar structure equations, while $\epsilon(h)$ and $p(h)$ do.
Fortunately, however, the adiabatic index $\Gamma(h)$ determines both
$\epsilon(h)$ and $p(h)$.  The definitions of the enthalpy in equation
(\ref{e:hdef}), and the adiabatic index in equation (\ref{e:Gammadef}),
can be written as the following pair of ordinary differential
equations for $\epsilon(h)$ and $p(h)$:
\begin{eqnarray}
\frac{dp}{dh}&=& \epsilon+p,\\
\frac{d\epsilon}{dh} &=& \frac{(\epsilon+p)^2}{p\,\Gamma(h)}.
\end{eqnarray}
Given an adiabatic index, $\Gamma(h)$, this first-order system of
equations can be integrated exactly (cf. Lindblom~\cite{Lindblom10})
to determine both $p(h)$ and $\epsilon(h)$:
\begin{eqnarray}
p(h)&=& p_0 \exp\left[\int_{h_0}^h \frac{e^{h'}dh'}{\mu(h')}\right],
\label{e:pintegral}\\
\epsilon(h)&=& p(h)\frac{e^h-\mu(h)}{\mu(h)},
\label{e:epsilonintegral}\\
\mu(h)&=& \frac{p_0e^{h_0}}{\epsilon_0+p_0}+
\int_{h_0}^h\frac{\Gamma(h')-1}{\Gamma(h')}e^{h'}dh',
\label{e:muintegral}
\end{eqnarray}
where $h_0$ is the lower bound on the range of enthalpies over which
the spectral expansion is performed, $\epsilon_0=\epsilon(h_0)$, and
$p_0=p(h_0)$.  While the integrals that appear in equations
(\ref{e:pintegral})--(\ref{e:muintegral}) can not generally be done
analytically, these integrals can nevertheless be done numerically
very efficiently.  I have found that they can be evaluated numerically
with double precision accuracy using Gaussian quadrature using only
ten or twelve collocation points.

\begin{figure}[h]
\includegraphics[width=6.5in]{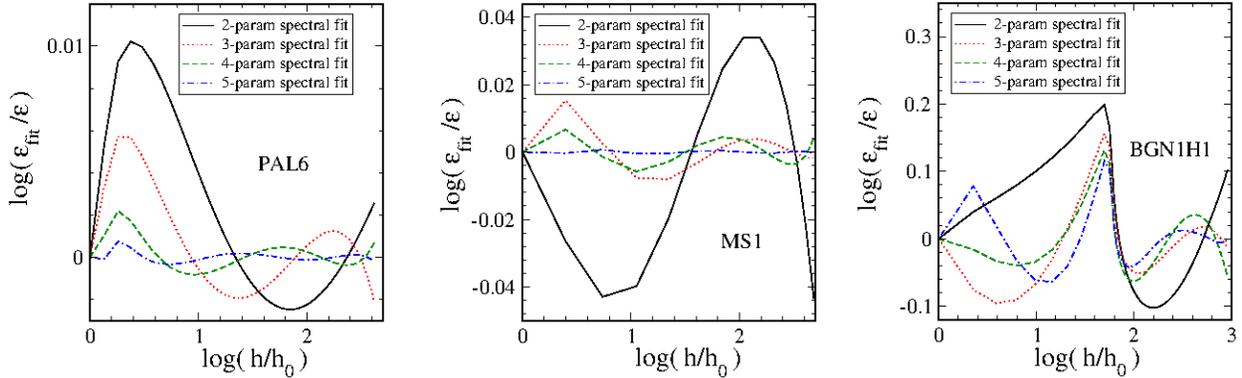}
\caption{Accuracies of the spectral representations of three of the
theoretical neutron-star equations of state.  These three
cases represent the equations of state with spectral representations
having the best (PAL6), 
the average (MS1), and the worst (BGN1H1) accuracies. \label{f:Fig6}}
\end{figure}

All the equations of state, $\epsilon(h,\gamma_{\,k})$ and
$p(h,\gamma_{\,k})$, defined by equations
(\ref{e:pintegral})--(\ref{e:muintegral}), using the adiabatic index
expansion given in equation (\ref{e:Gammadef}) are faithful for any
choice of basis functions.  To determine whether expansions of this
type are efficient and practical, I have used them to construct
spectral representations of 34 theoretical neutron-star equations of
state.  These theoretical equation of state models are cataloged in
Read, et al.~\cite{Read:2008iy}, and references are given there to the
original papers on which they are based.  These theoretical equations
of state are given as tables $\{p_i,\epsilon_i\}$ for
$i=1,...,N_{EOS}$, from which a set of corresponding $h_i$ can be
evaluated using equation (\ref{e:hdef}), once a suitable interpolation
formula is adopted.  These data were fit to these parametric equations
of state, $\epsilon=\epsilon(h,\gamma_{\,k})$ and
$p=p(h,\gamma_{\,k})$, by minimizing the quantity
$\Delta^{EOS}_{N_{\gamma_{\,k}}}$, defined by
\begin{eqnarray}
\left(\Delta^{EOS}_{N_{\gamma_{\,k}}}\right)^2 &=& \frac{1}{N_{\,{EOS}}}
\sum_{i=1}^{N_{\,{EOS}}}
\left\{
\log\left[\frac{\epsilon(h_{\,i},\gamma_{\,k})}{\epsilon_i}\right]\right\}^2,
\nonumber\end{eqnarray}
with respect to each of the spectral parameters $\gamma_{\,k}$.  Such
fits were performed using several choices for the spectral basis
functions $\Phi_k(h)$.  The following simple choice was found to be
quite effective for a wide range of these theoretical equations of
state,
\begin{eqnarray}
\Gamma(h) 
=\exp\left\{\sum_{k=0}^{N_{\gamma_{\,k}}-1} \gamma_{\,k} 
\left[\log\left(\frac{h}{h_0}\right)\right]^{k}\right\}.
\label{e:Gammapowerlaw}
\end{eqnarray}
This simple power-law expansion has therefore been adopted as the
standard to use for enthalpy-based representations of the equation of
state (cf. Lindblom~\cite{Lindblom10}).  The results of these fits are
summarized in Table~\ref{t:Table1}.  These results show that only two
or three spectral coefficients $\gamma_{\,k}$ are needed to produce
representations of these 34 theoretical neutron-star equations of
state with average errors of only a few percent errors over the entire
range of densities that occur in the cores of neutron stars.
\begin{table}[!ht]
\begin{tabular}{l|llll}
\hline
EOS & $\,\,\,\Delta^{EOS}_2$ &  $\,\,\,\Delta^{EOS}_3$ & $\,\,\,\Delta^{EOS}_4$ 
& $\,\,\,\Delta^{EOS}_5$ \\
\hline
PAL6 & $0.0032$ & $0.0016$ & $0.0005$ & $0.0002$ \\
MS1 & $0.0277$  & $0.0055$ &  $0.0035$ & $0.0003$ \\
BGN1H1 &  $0.0868$ &  $0.0495$ &  $0.0439$ & $0.0403$ \\
\hline
Average & $0.0323$ &  $0.0166$ & $0.0124$ & $0.0089$ \\
\hline
\end{tabular}
\caption{Summaries of the average accuracies of the direct spectral
  fits to 34 theoretical equations of state. The examples included in
  this table are the best case, PAL6, a typical average case, MS1, and
  the worst case, BGN1H1.  Also listed are the average errors for the
  spectral fits to all 34 theoretical equations of state.
  \label{t:Table1}}
\end{table}
Figure~\ref{f:Fig6} illustrates in more detail the errors in these
spectral equations of state for three particular cases: PAL6, MS1 and
BGN1H1.  These three represent the equations of state having the best
(PAL6), a typical average (MS1), and the worst (BGN1H1) spectral
representations.  These results show that even the worst case, the
BGN1H1 equation of state which has a strong phase transition within
the neutron-star density range, can be represented fairly accurately
with a spectral expansion having only a few spectral
coefficients.

\section{Spectral Solution of the Inverse Stellar Structure Problem}

The standard Oppenheimer-Volkoff version of the relativistic stellar
structure equations, (\ref{e:OVmass}) and (\ref{e:OVpressure}), are
not ideal for numerical work.  The problem is that the right side of
the pressure equation vanishes as the surface of the star is
approached.  The leading order term in the solution for the pressure
$p(r)$ near the surface goes like $p\propto
(R-r)^{\Gamma_0/(\Gamma_0-1)}$, where $\Gamma_0$ is the surface value
of the adiabatic index.  Consequently it is very difficult to solve
$p(R)=0$ numerically to determine the location of the surface of the
star.  This problem can be avoided by switching to the relativistic
enthalpy, $h$ defined in equation (\ref{e:hdef}), as the fundamental
thermodynamic variable. In this case the structure equation
(\ref{e:OVpressure}) becomes
\begin{eqnarray}
\frac{dh}{dr} = 
\frac{1}{\epsilon+p}\frac{dp}{dr} =
- \frac{m+4\pi r^3 p(h)}{r(r-2m)}.
\label{e:OVh}
\end{eqnarray}
This equation implies that the leading order term in the solution for
the enthalpy $h(r)$ goes like $h\propto (R-r)$ near the surface.
Consequently it is much easier numerically to solve $h(R)=0$ to
determine the location of the surface of the star in this case.

Equation~(\ref{e:OVh}) implies that the enthalpy $h(r)$ is a 
monotonically decreasing function in any star.  
It is possible, therefore to transform the relativistic structure equations
one more time by using $h$ instead of $r$ as the independent variable:
\begin{eqnarray}
\frac{dr}{dh} &=& 
- \frac{r(r-2m)}{m+4\pi r^3 p(h)},
\label{e:OVHr}\\
\frac{dm}{dh} &=& -\frac{4\pi\epsilon(h) r^3(r-2m)}
{m+4\pi r^3 p(h)}.
\label{e:OVHm}
\end{eqnarray}
The boundary conditions for these equations at the center of the star
where $h=h_c$ are $r(h_c)=m(h_c)=0$.  The solution to these equations is a
relativistic stellar model with the mass $m(h)$ and radius $r(h)$
expressed as functions of the enthalpy.  This version of the equations
has several advantages for numerical work.  First, the range of the
independent variable $h_c\geq h \geq 0$ is fixed before the
integration begins.  Therefore, it is not necessary to hunt for the
location of the surface by trying to solve $h(R)=0$ on the fly as part
of the solution process.  Second, the surface values of the total mass
$M$ and radius $R$ of the model are determined simply by evaluating
the solution at the endpoint of the integration where $h=0$: $M=m(0)$
and $R=r(0)$.  The disadvantage to this approach is that the standard
representation of the equation of state, $\epsilon=\epsilon(p)$ must
be transformed into a pair of functions $\epsilon=\epsilon(h)$ and
$p=p(h)$.  This can be done in a straightforward way numerically,
using the definition of the enthalpy in equation (\ref{e:hdef}).  In
practice this disadvantage is not a real problem.

These enthalpy based representations of the stellar structure
equations, (\ref{e:OVHr}) and (\ref{e:OVHm}), along with the enthalpy
based spectral representations of the equation of state, equations
(\ref{e:pintegral})--(\ref{e:Gammapowerlaw}), can be solved to
evaluate the macroscopic observables $M(h_c,\gamma_{\,k})$ and
$R(h_c,\gamma_{\,k})$ accurately and efficiently.  These observables
depend on the central value of the enthalpy $h_c$ in these models (in
much the same way as the standard representations depend on the
central pressure $p_c$), along with the spectral parameters
$\gamma_{\,k}$ that specify a particular equation of state.

My collaborator Nathaniel Indik and I have used these techniques to
construct solutions to the relativistic inverse stellar structure
problem~\cite{Lindblom12,Lindblom13}.  We compute the spectral
coefficients $\gamma_{\,k}$ that determine the approximate solution to
the inverse stellar structure problem by minimizing the quantity
$\chi(h_c^i,\gamma_{\,k})$ that measures the differences between the
model values of the observables $R(h_c^i,\gamma_{\,k})$ and
$M(h_c^i,\gamma_{\,k})$ and the data points $\{M_i,R_i\}$ taken from
  the exact curve of observables:
\begin{eqnarray}
\chi^2(h_c^i,\gamma_{\,k}) &=&
\frac{1}{N_{\,\mathrm{stars}}}\sum_{i=1}^{N_{\,\mathrm{stars}}}
\left\{\left[\log\left(\frac{M(h_{c}^{\,i},\gamma_{\,k})}{M_i}\right)\right]^2
+\left[\log\left(\frac{R(h_{c}^{\,i},\gamma_{\,k})}{R_i}\right)\right]^2\right\}.
\label{e:hchi}
\end{eqnarray}
The quantity $\chi(h_c^i,\gamma_{\,k})$ is minimized with respect to
variations in the $\gamma_{\,k}$, as well as the parameters $h_c^i$
that specify the central values of the enthalpies in the stars with
observables $\{R_i,M_i\}$.  This minimization is accomplished using
standard numerical methods.

\begin{figure}[!ht]
\includegraphics[width=2.0in]{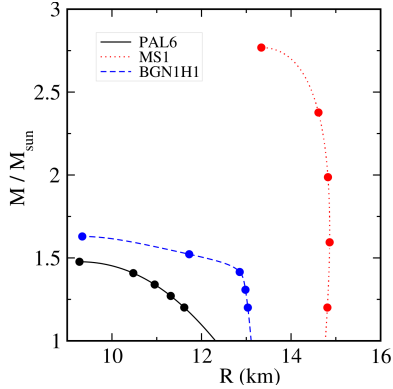}
\caption{Mock data for the masses and radii $\{R_i,M_i\}$ were
  selected from the mass-radius curves of each of the 34 theoretical
  neutron-star equations of state.  Shown here are the mass-radius
  curves for the PAL6, MS1 and the BGN1H1 equations of state with the
  $N_\mathrm{stars}=5$ data points selected from each curve.  Mock
  data were chosen to be equally spaced in mass between $1.2M_\odot$
  (a typical minimum astrophysical neutron-star mass) and the maximum
  mass model for each particular equation of state.  \label{f:Fig7}}
\end{figure}

We have tested this method by constructing sets of mock observables
$\{R_i,M_i\}$ based on the 34 theoretical neutron star equations of
state described earlier.  For each equation of state, we select $2\leq
N_\mathrm{stars}\leq 5$ pairs of observables.  The masses of these
mock data are chosen to be uniformly spaced between $1.2M_\odot$ (a
typical minimum astrophysical neutron-star mass), and the maximum-mass
model associated with that particular equation of state.
Figure~\ref{f:Fig7} illustrates these mock data points for the
$N_\mathrm{stars}=5$ cases of the PAL6, the MS1 and the BGN1H1
equations of state.

We then solve the inverse stellar structure problem by solving for the
parameters $\gamma_{\,k}$ and $h_c^i$ that minimize
$\chi(h_c^i,\gamma_{\,k})$ for the data sets consisting of
$N_\mathrm{stars}$ observable data pairs from each of the 34
theoretical equations of state.  We tested the accuracy of the
equations of state determined in this way by comparing the resulting
spectral model equation of state $\epsilon(h,\gamma_{\,k})$ with the
tabulated exact theoretical equation of state $\{p_i,\epsilon_i,h_i\}$
that was used to construct the mock data.  We evaluate the differences
between these using the average error measure
$\Delta^{MR}_{N_{\gamma_{\,k}}}$ defined by
\begin{eqnarray}
\left(\Delta^{MR}_{N_{\gamma_{\,k}}}\right)^2 &=&
\frac{1}{N_{\,\mathrm{EOS}}} \sum_{i=1}^{N_{\,\mathrm{EOS}}} \left\{
\log\left[\frac{\epsilon(h_{\,i},\gamma_{\,k})}{\epsilon_i}\right]\right\}^2.
\end{eqnarray}
These results are summarized in Table~\ref{t:Table2}.  The average
errors in the spectral equations of state determined by solving the
inverse stellar structure problem $\Delta^{MR}_{N_{\gamma_{\,k}}}$ are
given in the left columns of Table~\ref{t:Table2}.  These error
values, $\Delta^{MR}_{N_{\gamma_{\,k}}}$, can be compared to those
constructed earlier using direct spectral fits to the equation of
state tables, $\Delta^{EOS}_{N_{\gamma_{\,k}}}$, which are reproduced
in the right columns of Table~\ref{t:Table2}.  We see that the
accuracies of the equations of state determined by solving the inverse
stellar structure problem using spectral methods are somewhat less
accurate than the optimal spectral fits to those equations of state.
This is not surprising.  For the smoothest equations of state however,
like PAL6, the difference between the inverse stellar structure
equation of state and the ones determined by doing direct spectral
fits are negligible.  Even for equations of state having sharp phase
transitions, like BGN1H1, the errors in the inverse stellar structure
equations of state are only a few times worse than the best fits.
\begin{table}[!hb]
\begin{tabular}{l|llll|llll}
\hline
EOS & $\,\,\,\Delta^{MR}_2$ &  $\,\,\,\Delta^{MR}_3$ & $\,\,\,\Delta^{MR}_4$ 
& $\,\,\,\Delta^{MR}_5$ 
& $\,\,\,\Delta^{EOS}_2$ &  $\,\,\,\Delta^{EOS}_3$ & $\,\,\,\Delta^{EOS}_4$ 
& $\,\,\,\Delta^{EOS}_5$ \\
\hline
PAL6 & $0.0034$ & $0.0018$ & $0.0007$ & $0.0003$ 
& $0.0032$ & $0.0016$ & $0.0005$ & $0.0002$ \\
MS1 & $0.0474$  & $0.0157$ &  $0.0132$ & $0.0009$
& $0.0277$  & $0.0055$ &  $0.0035$ & $0.0003$ \\
BGN1H1 &  $0.1352$ &  $0.1702$ &  $0.1356$ & $0.1382$
&  $0.0868$ &  $0.0495$ &  $0.0439$ & $0.0403$ \\
\hline
Average & $0.0396$ &  $0.0289$ & $0.0276$ & $0.0239$
& $0.0323$ &  $0.0166$ & $0.0124$ & $0.0089$ \\
\hline
\end{tabular}
\caption{Accuracies of the equations of state determined by the
  spectral approach to the inverse stellar structure problem based on
  mass-radius data. The examples included in this table are the best
  case, PAL6, a typical median case, MS1, and the worst case, BGN1H1.
  Also listed are the average errors for all 34 theoretical equations
  of state.  The accuracies from Table~\ref{t:Table1} of the direct
  spectral fits to these equations of state are also listed in the
  right columns of this table for comparison.
\label{t:Table2}}
\end{table}

\begin{figure}[!ht]
\includegraphics[width=6.5in]{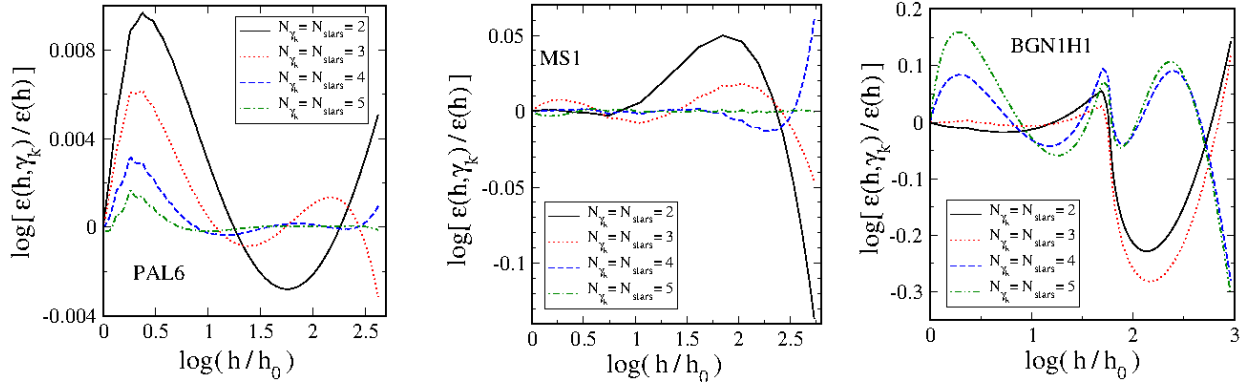}
\caption{Accuracies of the spectral representations determined by
  solving the inverse stellar structure problem using mock mass-radius
  data from three of the theoretical neutron-star equations of state.
  These three cases represent the equations of state whose spectral
  representations have the best (PAL6), the average (MS1), and the
  worst (BGN1H1) accuracies.
\label{f:Fig8}}
\end{figure}

The overall accuracies of the equations of state determined by solving
the spectral version of the inverse stellar structure problem are
quite impressive.  These equations of state can be determined using
spectral methods with average errors of just a few percent over the
entire range of neutron star densities, using (high quality)
measurements of the masses and radii of just two or three neutron
stars.  Even the worst case, an equation of state with a sharp phase
transition in the neutron-star density range, is determined with
accuracy levels around 15\% using high quality measurements of just a
few masses and radii.  Figure~\ref{f:Fig8} illustrates in more detail
the errors in the equations of state obtained with mock mass-radius
data from the three equations of state PAL6, MS1 and BGN1H1.  These
error graphs are qualitatively similar to those displayed in Fig.~\ref{f:Fig6}
for the direct spectral fits.

\section{Inverse Stellar Structure Problem Using Tidal Deformabilities}

The discussion of the inverse stellar structure problem in this talk
has focused up to this point on developing a nuclear-physics model
independent method for using mass-radius data to determine the
equation of state of neutron star matter.  The advances in
observational methods over the past few years has dramatically
increased the number and quality of the observations of neutron star
masses and radii.  It seems likely that these improvements will
continue, and that these observations will eventually provide high
enough quality data to determine the equation of state with good
accuracy.  Other types of observations can provide measurements of
other neutron-star observables, and in many cases these observables
can also be used to solve the inverse stellar structure problem.  In
particular Flannigan and Hinderer~\cite{Hinderer2008a} have pointed
out (and this has been confirmed by a number of others, e.g., Lackey,
et al., \cite{Lackey2012, Lackey2013}) that the masses and tidal
deformabilities of neutron stars could be measured by observing the
gravitational waveforms produced by the inspiral and mergers of
neutron-star binary systems.  In this last part of my talk I will
describe some recent work (done in collaboration with Nathaniel Indik)
that tests how well the spectral approach to the inverse stellar
structure problem works when applied to (ideal) mass and tidal
deformability data.

The tidal deformability, $\Lambda$, measures the amount by which a
star distorts it shape in response to the application of a tidal force
from a nearby star (or black hole).  The tidal deformability depends
on the equation of state of the star, and therefore its measurement
can in principle be used to determine the equation of state of the
stellar matter.  For a given equation of state, the tidal
deformability can be calculated for model stars using methods first
developed by Hinderer~\cite{Hinderer2008, Hinderer2009}, and then
transformed into the somewhat more efficient method described here by
Lindblom and Indik~\cite{Lindblom13}.  The stellar structure
equations, (\ref{e:OVHr}) and (\ref{e:OVHm}), are solved for $m(h)$
and $r(h)$, together with a third equation that determines a function
$y(h)$ which describes the time independent quadrapole deformations of
a relativistic star,
\begin{eqnarray}
\frac{dy}{dh}&=&
\frac{(r-2m)(y+1)y}{m+4\pi r^3p}+y
+\frac{(m-4\pi r^3\epsilon)y}{m+4\pi r^3 p}
+\frac{4\pi r^3 (5\epsilon+9p)-6r}{m+4\pi r^3 p}
+\frac{4\pi r^3(\epsilon+p)^2}{p\Gamma(m+4\pi r^3 p)}
-\frac{4(m+4\pi r^3 p)}{r-2m}.\qquad
\label{e:dydhEquation}
\end{eqnarray}
The appropriate initial condition for $y(h)$ at the center of the
star, $h=h_c$, is $y(h_c)=2$.  After this integration, the values of
the resulting functions are evaluated at the surface of the star,
$h=0$, to determine the quantities $M=m(0)$, $R=r(0)$, and $Y=y(0)$.
The (dimensionless) tidal deformability $\Lambda$ is then determined
by the expression,
\begin{eqnarray}
&&\!\!\!\!\!
\Lambda(C,Y)=
\frac{16}{15 \Xi}(1-2C)^2[2+2C(Y-1)-Y],\qquad
\label{e:DefTidalDeformationAppendix}
\end{eqnarray}
where $C=M/R$, and $\Xi$ is given by
\begin{eqnarray}
\Xi(C,Y)&=&4C^3[13-11Y+C(3Y-2)+2C^2(1+Y)]
+3(1-2C)^2[2-Y+2C(Y-1)]\log(1-2C)\nonumber\\
&&+2C[6-3Y+3C(5Y-8)].
\end{eqnarray}

For a given equation of state, the mass $M(h_c,\gamma_{\,k})$ and
tidal deformability $\Lambda(h_c,\gamma_{\,k})$ of a stellar model
will depend on the central value of the enthalpy $h_c$ as well as the
parameters $\gamma_{\,k}$ that characterized the equation of state.
Using the same spectral approach to the inverse stellar structure
problem described earlier, these parameters can be fixed by matching
the model observables, $\{\Lambda(h_c^i,\gamma_{\,k}),
M(h_c^i,\gamma_{\,k})\}$, with data $\{\Lambda_i,M_i\}$ from
the exact curve of masses and tidal deformabilities.  As before, this
matching is done by minimizing the quantity, $\chi(h_c^i,\gamma_{\,k})$,
defined by
\begin{eqnarray}
\chi^2(h_c^i,\gamma_{\,k}) &=&
\frac{1}{N_{\,\mathrm{stars}}}\sum_{i=1}^{N_{\,\mathrm{stars}}}
\left\{\left[\log\left(\frac{M(h_{c}^{\,i},\gamma_{\,k})}{M_i}\right)\right]^2
+\left[\log\left(\frac{\Lambda(h_{c}^{\,i},\gamma_{\,k})}
{\Lambda_i}\right)\right]^2\right\},
\label{e:hchi2}
\end{eqnarray}
with respect to variations in the parameters $h_c^i$ and
$\gamma_{\,k}$.  The resulting values of $\gamma_{\,k}$ determine an
equation of state that serves as an approximate solution to the
inverse stellar structure problem.  The results of our solutions to
this problem are described in Table~\ref{t:Table3} for mock
$\{\Lambda_i,M_i\}$ data computed with the 34 theoretical neutron-star
equations of state described earlier.  The accuracies of these
solutions are determined by evaluating the equation of state error
measures $\Delta^{M\Lambda}_{N_{\gamma_{\,k}}}$, defined by
\begin{eqnarray}
\left(\Delta^{M\Lambda}_{N_{\gamma_{\,k}}}\right)^2 &=&
\frac{1}{N_{\,\mathrm{EOS}}} \sum_{i=1}^{N_{\,\mathrm{EOS}}} \left\{
\log\left[\frac{\epsilon(h_{\,i},\gamma_{\,k})}{\epsilon_i}\right]\right\}^2.
\end{eqnarray}
Also listed in Table~\ref{t:Table3} are the analogous error measures
$\Delta^{MR}_{N_{\gamma_{\,k}}}$ for the equations of state determined
by solving the inverse stellar structure problem using mass and radius
data instead of mass and tidal deformabilities.  These results show
that mass and tidal deformability data is about as effective as
mass and radius data in determining the equation of state.
\begin{table}[!ht]
\begin{tabular}{l|llll|llll}
\hline
EOS & $\,\,\,\Delta^{M\Lambda}_2$ &  $\,\,\,\Delta^{M\Lambda }_3$ 
& $\,\,\,\Delta^{M\Lambda}_4$ 
& $\,\,\,\Delta^{M\Lambda}_5$ 
& $\,\,\,\Delta^{MR}_2$ &  $\,\,\,\Delta^{MR}_3$ & $\,\,\,\Delta^{MR}_4$ 
& $\,\,\,\Delta^{MR}_5$ \\
\hline
PAL6 
& $0.0034$ & $0.0019$ & $0.0008$ & $0.0003$ 
& $0.0034$ & $0.0018$ & $0.0007$ & $0.0003$ \\
MS1 
& $0.0465$  & $0.0141$ &  $0.0129$ & $0.0008$ 
& $0.0474$  & $0.0157$ &  $0.0132$ & $0.0009$\\
BGN1H1 
&  $0.1356$ &  $0.1652$ &  $0.1445$ & $0.1363$ 
&  $0.1352$ &  $0.1702$ &  $0.1356$ & $0.1382$\\
\hline
Average 
& $0.0403$ &  $0.0295$ & $0.0304$ & $0.0291$ 
& $0.0396$ &  $0.0289$ & $0.0276$ & $0.0239$\\
\hline
\end{tabular}
\caption{Accuracies of the equations of state determined by the
  spectral approach to the inverse stellar structure problem based on
  mass-tidal deformability data. The examples included in this table
  are the best case, PAL6, a typical median case, MS1, and the worst
  case, BGN1H1.  Also listed are the average errors for all 34
  theoretical equations of state.  The accuracies from
  Table~\ref{t:Table2} of the solutions based on mass-radius data are
  also listed in the right columns of this table for comparison.
\label{t:Table3}}
\end{table}

\section{Summary}

This talk has discussed various aspects of the relativistic inverse
stellar structure problem, i.e.  the problem of determining the
equation of state of the stellar matter from a knowledge of a curve
through the space of macroscopic observables of those stars.  Some
recent work on the use of spectral methods was described.  Spectral
methods are a very efficient way to represent functions, and the
particular spectral expansions described here were shown to be capable
of describing a wide variety of theoretical neutron-star equations of
state with accuracies of a few percent using spectral approximations
with only two or three spectral parameters.  It was also shown that
these spectral parameters could be determined by fitting models of the
masses and radii or the masses and tidal deformabilities of neutron
stars with accuracies that are only somewhat less accurate that the
optimal spectral fits to those equations of state.

The spectral solutions to the inverse stellar structure problem
described here were based on the use of very high quality measurements
of the macroscopic observables that are uniformly distributed across
the astrophysically relevant range of neutron-star masses.  It remains
to be seen how well these spectral methods perform when more realistic
data us used, i.e. data with finite measurement errors and data that
is maldistributed in some way.  I plan to investigate those issues
at some point in the near future.


\begin{theacknowledgments}
I thank the organizers of the Fifth Leopoldo Garc\'ia-Col\'in Mexican
Meeting on Mathematical and Experimental Physics for arranging a
scientifically interesting and extremely enjoyable conference.  I am
grateful for having had the opportunity to participate, to renew old
scientific friendships, and to make a number of new ones.  I also
thank the Mathematical Sciences Center at Tsinghua University in
Beijing, China for their hospitality during the time this manuscript
was being written.  The research reported in this talk was supported
by a grant from the Sherman Fairchild Foundation and by NSF grants
PHY1005655 and DMS1065438.
\end{theacknowledgments}
\vfill\break


\bibliographystyle{aipproc}   

\bibliography{./References.bib}

\IfFileExists{\jobname.bbl}{}
 {\typeout{}
  \typeout{******************************************}
  \typeout{** Please run "bibtex \jobname" to optain}
  \typeout{** the bibliography and then re-run LaTeX}
  \typeout{** twice to fix the references!}
  \typeout{******************************************}
  \typeout{}
 }

\end{document}